\documentclass[lettersize,journal]{IEEEtran}
\usepackage{amsmath,amsfonts}
\usepackage{algorithm}
\usepackage{array}
\usepackage{textcomp}
\usepackage{stfloats}
\usepackage{url}
\usepackage{verbatim}
\usepackage{graphicx}
\usepackage{cite}

\usepackage{ifpdf}
\usepackage{hhline}
\usepackage{float}
\usepackage{booktabs}
\usepackage{enumerate}
\usepackage{color}
\usepackage{amsthm,amssymb}
\usepackage{caption}
\usepackage{algpseudocode}

\usepackage{bm,textcomp}
\usepackage{mathrsfs}
\usepackage[normalem]{ulem}
\useunder{\uline}{\ul}{}
\usepackage{multirow}

\usepackage{tikz,pgf}
\usepackage{threeparttable}
\usetikzlibrary{shapes.gates.logic.US,trees,shadows,positioning,arrows}
\usepackage{ragged2e}

\usepackage{tabularx}
\usepackage{bbding,utfsym}
\usepackage{diagbox}
\usepackage[english]{babel}
\usepackage{makecell}
\usepackage{hyperref}
\hypersetup{hidelinks}
\usepackage{arydshln} 

\renewcommand{\algorithmicrequire}{\textbf{Input:}}
\renewcommand{\algorithmicensure}{\textbf{Output:}}

\begin{document}

\title{Performance-lossless Black-box Model Watermarking}

\author{Na Zhao, Kejiang Chen\IEEEauthorrefmark{1}, Weiming Zhang\IEEEauthorrefmark{1}, and Nenghai Yu
\thanks{The authors are with CAS Key Laboratory of Electro-Magnetic Space Information, University of Science and Technology of China, Hefei 230026, China and Anhui Province Key Laboratory of Cyberspace Security Situation Awareness and Evaluation. (e-mail: znzhaona@mail.ustc.edu.cn; chenkj@ustc.edu.cn; zhangwm@ustc.edu.cn; ynh@ustc.edu.cn.)}
\thanks{Corresponding authors: Kejiang Chen and Weiming Zhang (Email:chenkj@ustc.edu.cn, zhangwm@ustc.edu.cn).}
\thanks{Our code will be public soon: \href{https://github.com/Othir/BranchWM/}
{https://github.com/Othir/BranchWM/}.}}

\markboth{Submitted to IEEE TDSC}%
{Zhao \MakeLowercase{\textit{et al.}}: Performance-lossless Black-box Model Watermarking}

\IEEEpubid{0000--0000/00\$00.00~\copyright~2024 IEEE}

\IEEEpubidadjcol

\maketitle

\begin{abstract}
With the development of deep learning, high-value and high-cost models have become valuable assets, and related intellectual property protection technologies have become a hot topic. However, existing model watermarking work in black-box scenarios mainly originates from training-based backdoor methods, which probably degrade primary task performance. To address this, we propose a branch backdoor-based model watermarking protocol to protect model intellectual property, where a construction based on a message authentication scheme is adopted as the branch indicator after a comparative analysis with secure cryptographic technologies primitives. We prove the lossless performance of the protocol by reduction. In addition, we analyze the potential threats to the protocol and provide a secure and feasible watermarking instance for language models. 
\end{abstract}

\begin{IEEEkeywords}
Performance-lossless, black-box, watermark, branch.
\end{IEEEkeywords}

\section{Introduction}
\IEEEPARstart{R}{ecently}, the development of deep learning technology, represented by GPTs~\cite{radford2018gpt,radford2019gpt2,brown2020gpt3,openai2023gpt4}, GANs~\cite{goodfellow2014GAN,brock2019BigGAN,karras2021StyleGAN}, Diffusion models~\cite{dhariwal2021diffusion,ho2022videoDM,rombach2022SDM}, etc., has become increasingly mature. Its applications~\cite{chatgpt,bard,claude} in Computer Vision (CV), Natural Language Processing (NLP), etc., have demonstrated tremendous commercial value. However, developing high-performance models entails substantial costs, including the involvement of machine learning experts with rich experience, extensive and high-quality task datasets, and expensive computational resources. Development teams maintain legitimate intellectual property rights over their produced models and often refrain from open-sourcing critical model details. They provide on-demand model functionality services to users through a commercial paradigm of remote access to model input-output Application Programming Interfaces (APIs), such as OpenAI's ChatGPT~\cite{chatgpt}, Google's Bard~\cite{bard}, and Anthropic's Claude~\cite{claude}. 

While development teams take great care in deploying these valuable model assets, they cannot completely prevent malicious users from disregarding intellectual property. We consider a threat scenario wherein model service providers (model development teams) possessing valuable model assets offer business arrangements allowing third-party application developers (application service providers) to access model functionality via black-box APIs. However, dishonest application developers may attempt to infringe on the intellectual property of model service providers. They do so by leveraging network access restrictions between different countries to resell API access rights to users, using low-quality models to impersonate high-performance models when deploying vertical applications such as chatbots to deceive uninformed application users, or even directly deploying the unlicensed model in the name of a self-developed high-performance model. Such behavior obtaining improper benefits has seriously damaged the intellectual property rights, reputation, and economic interests of the model development team. Security measures need to be taken to detect and confirm the legitimacy of the model.

Watermarking serves as an effective solution for protecting intellectual property. Existing model watermarking works can generally be divided into three types based on the model knowledge needed for copyright verification or the ability to control the model: white-box watermarking, black-box watermarking, and no-box watermarking. White-box verification watermarking assumes that the model is completely transparent, allowing the verifier to manipulate the model within the verification rules to detect the watermark signal. In black-box verification watermarking scenarios, the verifier lacks knowledge of the inner workings of the model but can manipulate the inputs of the model to obtain specific outputs, thereby detecting the watermark signal. In the case of no-box validation watermarking, the verifier only has access to the model's output to detect the watermark signal. 
In the black-box threat scenario, suspect models are difficult to obtain, which limits white-box model watermarking methods. No-box model watermarking requires independently meaningful model outputs, which makes it useful only for generative tasks. Therefore, black-box model watermarking is more practical. 

\IEEEpubidadjcol 

Black-box model watermarking works mainly rely on backdoor techniques. In 2018, Adi et al.~\cite{adi2018turning} first established the connection between backdoor models and model watermarking, introducing the first backdoor-based model watermarking work. Verifiers activate a pre-embedded backdoor in the model by inputting trigger samples during the inference process and detecting watermark signals in the output for model ownership confirmation. Existing backdoor-based model watermarking works~\cite{darvish2019deepsigns,le2020adversarial,adi2018turning,chen2019blackmarks,li2022move,jia2022subnetwork,lounici2021yes,lao2022identification,jia2021entangled,guo2018watermarking,zhu2020secure,maung2021piracy,li2022untargeted,li2019prove,lao2022deepauth,yang2021robust,charette2022cosine,namba2019robust,bansal2022certified,zhong2020protecting,mehta2022aime} usually involve fine-tuning or retraining the model's weights to embed the backdoor, which essentially constitutes multitask learning. Due to the incomprehensible nature of neural networks, despite attempts to separate the influences between the two tasks through carefully designed loss functions~\cite{jia2021entangled}, weight selections~\cite{lao2022identification,namba2019robust,bansal2022certified}, trigger sets~\cite{darvish2019deepsigns,adi2018turning,zhang2018protecting,jia2022subnetwork,lounici2021yes,lao2022identification,jia2021entangled,guo2018watermarking,zhu2020secure,maung2021piracy,zhong2020protecting}, etc., there is no reliable theoretical analysis to prove that the watermarking task will not distort the primary task. Their experimental results also indicate that embedding watermarks impacts the accuracy of image classification models. 

An ideal model watermark should not affect the model’s original functionality. To meet this requirement, we adopt the idea proposed by Adi et al.~\cite{adi2018turning} that watermarks can be transformed from backdoors and draw inspiration from the undetectable black-box construction in the work of Goldwasser et al.~\cite{goldwasser2022planting}. We propose a provably performance-lossless black-box model watermarking protocol named BranchWM. Specifically, we add a parallel branch for forensic tasks outside the original model structure, decoupling the original task of the model from the watermark forensic task. Then, we construct a trigger for forensic tasks using an existence unforgeable Message Authentication Code (MAC) scheme (a secure cryptographic primitive). 
Without owners' keys and white-box access to the model, it is hard for attackers to know trigger patterns and forge triggers. Therefore, the probability that a normal sample triggers the forensic branch is computationally ignored, which leads to a performance-loss model watermarking. 
Finally, we provide an instantiation algorithm for our proposed watermark using language models and analyze potential interference attacks during actual forensic processes as well as improved watermark instances.

Our major contributions can be summarized as follows:
\begin{enumerate}
    \item We first propose a provably performance-lossless black-box model watermarking protocol named BranchWM, which is based on the branch structure and secure cryptography.
    \item We employ an MAC with existential unforgeability to construct a forensic trigger named Mac-based Unaccessible and Fast Trigger (MUFT), which is both unaccessible and can be quickly detected.
    \item We analyze the potential attacks in the forensic process and provide a secure and feasible watermarking instance for language models.
\end{enumerate}

\section{Related Work}
\subsection{Backdoor Attacks}
Backdoor attacks aim to embed hidden backdoors into deep neural networks, allowing the model to perform normally on regular samples but activate the internal backdoor when exposed to predefined triggers, resulting in malicious or poor-quality predictions. 

Backdoor attacks based on poisoned data~\cite{gu2017badnets,turner2019label,chen2017blended,saha2020hidden,turner2018clean} have long been a topic of concern. When model developers face resource constraints, whether due to a lack of training data or limited computing power, they may purchase datasets from third parties or delegate training tasks to third parties, thereby exposing themselves to the risk of attackers injecting poisoned data into the training data. The attacker poisons the training data, causing the model to learn the knowledge of the poisoned data during the training process to plant the backdoor. Backdoor attacks based on poisoned training code~\cite{bagdasaryan2021blind} constitute a form of blind attack on unknown models. The attacker inserts malicious code into the model's training code, automatically poisons the data and controls the loss function, and injects new tasks into the model during the training process to plant backdoors.

Whether the work is based on poisoned data or poisoned training code, it usually involves injecting backdoors through training. The knowledge learned from this poisoned content often leads to poor predictions on specific triggers. When the knowledge is used as a signature of model developers, this backdoor technique can also be used for model copyright verification. 

\subsection{Watermarking for Models}
Model watermarking is a technology used to protect and track machine learning models, similar to the application of digital watermarking in digital media. The main goal of model watermarking is to embed a unique identifier or label in the model without affecting the performance or accuracy of the model, thereby identifying and verifying the legitimacy of the model. This helps prevent unauthorized copying, misuse or theft of the model while also serving as a means of proving the provenance and integrity of the model. In the context of the remote serving paradigm of models, we mainly focus on two types of model watermarking: black-box watermarking and no-box watermarking.

\subsubsection{Black-box Watermarking}
The ``black box'' is an image description of the verifier's knowledge of the model, meaning that the verifier's knowledge of the model is limited to queries and responses. Model owners use techniques such as backdoor attacks to embed signature-related watermark information in the model. They use specific trigger query-response pairs to detect watermarks for verification. 

Adi et al.~\cite{adi2018turning} introduced a relationship between malicious backdoor attacks and benign watermark defenses for the model, converting the watermarking task of deep neural networks to a backdoor task. Existing model watermarking approaches based on backdoor design share the same thought as methods such as training-based backdoors. Model owners use specific input-output pairs as watermark patterns, which is a form of knowledge that needs to be learned. This knowledge is injected by training the model. Then, during ownership disputes, specific input triggers are used to activate the watermark backdoor and obtain relevant specific outputs for copyright verification. However, these backdoor attacks rely on injecting the backdoor during training. The backdoor task and the model's original task are entangled during the training process, which makes it challenging to ensure that the original task performance is not distorted. The same goes for watermarking tasks. Ensuring that the watermark knowledge does not impact the original task performance of the model is a problem that remains to be addressed theoretically. 

\subsubsection{No-box Watermarking}
There is an ownership dispute scenario where the validator does not even have access to the ``box'' of the suspect model. The model owner embeds a special watermark identifier in the model's output. When ownership disputes arise, under the strict condition that there are only some unknown inputs to the model, they can extract copyright information from the output to complete verification of the source model. 

For tasks related to copyright protection of language generation models, methods such as synonym replacement-based no-box watermarking were proposed, and subsequent work further improved the impact of watermarking scheme on text quality~\cite{he2022cater,he2022protecting}. Kirchenbauer et al.~\cite{kirchenbauer2023watermark} proposed a zero-bit watermarking scheme that adds biased signals based on greenlists and redlists to the outputs (i.e., logits) of the model. While the paper targeted a binary classification traceability task for human-machine text, it can be applied to model copyright verification tasks in a no-box scenario. Some works~\cite{lee2023code,zhao2023provableRobust,fu2023semantic,liu2023private,yoo2023color} have extended and improved upon this work~\cite{kirchenbauer2023watermark}. Zhao et al.~\cite{zhao2023textCos} transformed the model watermarking work~\cite{charette2022cosine} designed for image classification tasks to language generation tasks in the context of no-box watermarking. While no-box model watermarking is limited to generation tasks, it can be used to enhance the concealment of backdoor-based watermarking.

\section{A Performance-lossless Branch Watermarking Protocol for Model}
In this section, we provide a comprehensive description of the threat model targeted in this paper. Then, we explain the motivation behind proposing a black-box performance-lossless branch backdoor-based watermarking for model. When introducing the proposed method, we first clarify the motivation and then provide a formal definition. We then present the core construction of the watermarking based on MAC and prove its performance-lossless property. Finally, to highlight the differences between our work and that of Goldwasser et al.~\cite{goldwasser2022planting}, we analyze the impact of three secure cryptographic primitives——MAC, hash, and digital signatures—on the construction of performance-lossless branch watermarking and explain the advantages of using MAC.

\begin{figure*}[t]
    \centering
    \includegraphics[width=0.98\linewidth]{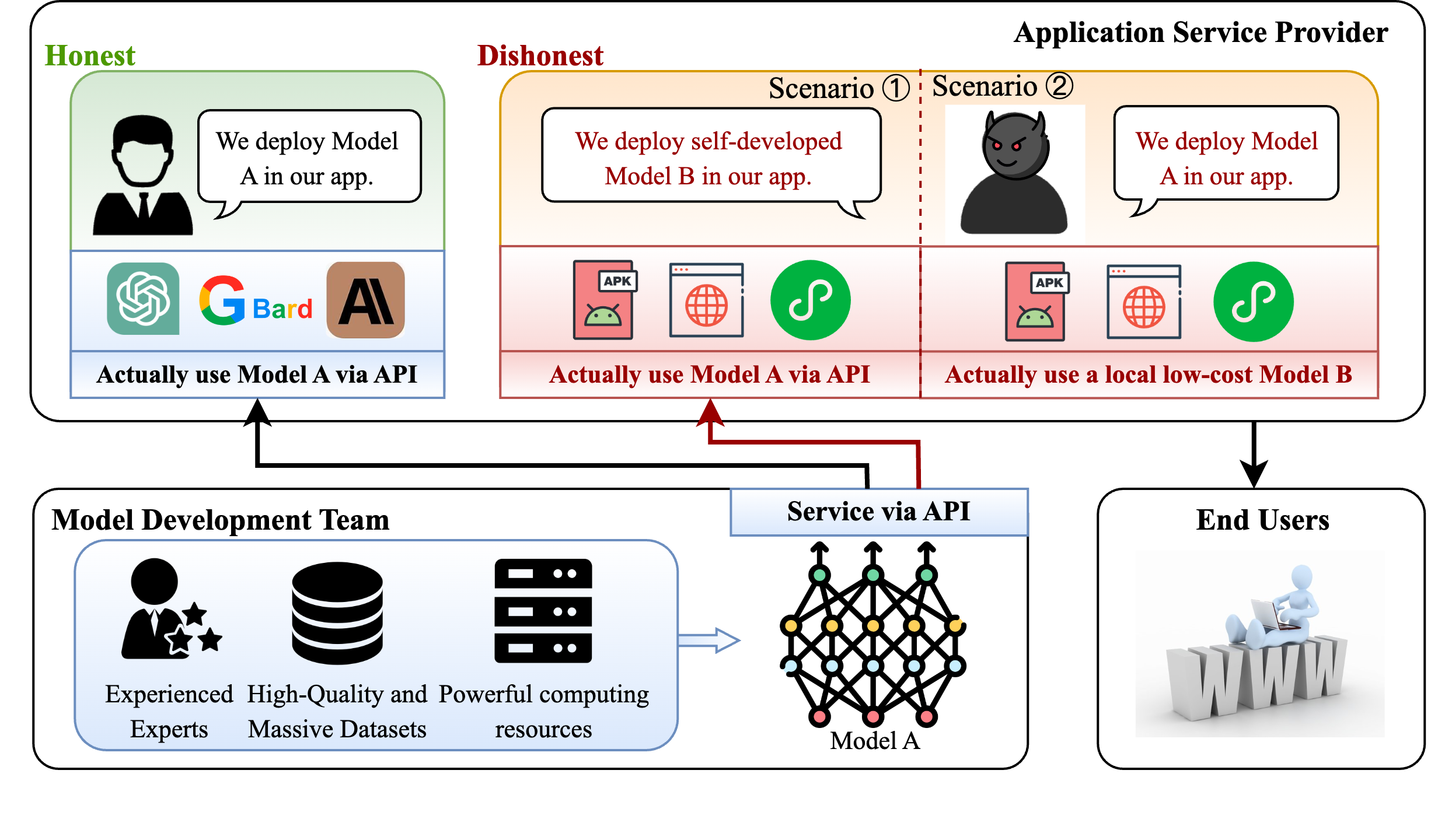}
    \caption{There are three roles in the threat scenario. The model development team provide black-box deployment to the application service provider. The latter may dishonestly provide application services to end users, which we summarize as two cases of dishonesty in the red zone.}
    \label{fig:threat}
\end{figure*}

\subsection{Threat Model}
In the threat model of the model remote service scenario, there are three key roles: the model development team, application service provider and end users, as shown in Figure~\ref{fig:threat}.

\textbf{The model development team} consists of machine learning experts. They invest considerable resources and effort into developing high-performance models, such as large language models. 
They act as model owners and provide remote access to model functionality via APIs to application service providers in exchange for a fee.

\textbf{The application service provider} is a team of non-machine learning experts who purchase model functionality services from the model owner. They leverage remote APIs to deploy specific applications (such as poe.com) and deliver well-encapsulated, user-friendly, and easy-to-use model applications to end-users.

\textbf{End users} are customers who purchase specific application functionalities from the application service provider. They can enjoy model services without having relevant expertise.

The threats considered here come from dishonest application service providers. Honest application service providers will strictly abide by the API usage agreement, while dishonest application service providers may abuse the model's remote API for unauthorized application development. Two situations are given here:
\begin{enumerate}
    \item The application provider claims to use a well-known high-performance model A via the remote API interface but in fact uses a low-cost model B. Their aim is to profit from users at low cost while damaging the reputation of the model in question.
    \item The application provider uses a well-known high-performance model A through the remote API but deliberately claims that it is self-developed model B. This allows them to gain market share and profits while compromising the intellectual property of model owners.
\end{enumerate}

In practical situations, the model owner and the application service provider may be the same team. However, this involves some internal personnel management issues that are not within the scope of this research. This paper treats the model owner and application service provider as two independent entities and primarily explores external threats to the model owner.

\subsection{Motivation}
Goldwasser et al.~\cite{goldwasser2022planting} defined the black-box undetectability of the backdoor model based on computational indistinguishability and constructed a black-box undetectable backdoor from a digital signature scheme that meets strong unforgeable security. The core is a branch structure that relies on secure cryptographic primitives.
Inspired by  the thought of conversion between backdoor and watermarking~\cite{adi2018turning}, we use the black-box undetectable backdoor to construct a black-box performance-loss watermarking for model. 

However, a digital signature is an asymmetric cryptography technology, leading to its low computational efficiency. If Goldwasser et al.'s work~\cite{goldwasser2022planting} is used directly for model watermarking, the running overhead is too heavy to deploy in real scenarios. Therefore, we considered more efficient cryptographic techniques with similar security properties—collision-resistant hash functions and existentially unforgeable message authentication codes—and ultimately chose the latter to construct the watermarking.

\subsection{Planting a Parallel Branch in Model API}
Remote services have the advantages of a black box model, allowing watermarking modules to be added to the API. We propose a branch backdoor-based watermarking protocol for model. 
The protocol consists of two modules, the model module and the watermarking module, with the watermarking module further divided into the inner module and the outer module. The outer module includes a trigger generator and evidence verifier submodules. 
The inner module includes a trigger detector and evidence generator submodules deployed within the API.

\subsubsection{Overview}
\begin{figure}
    \centering
    \includegraphics[width=1\linewidth]{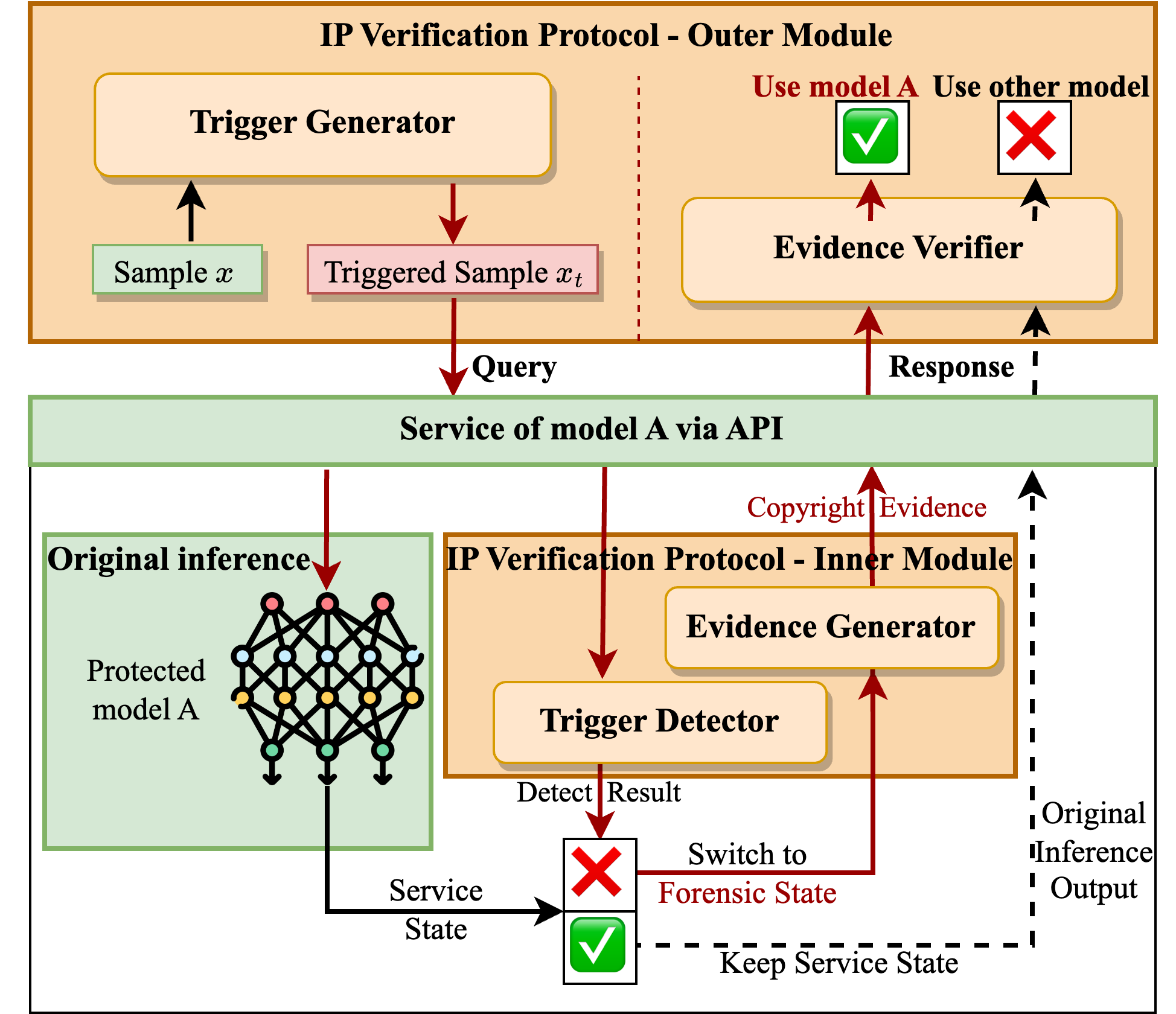}
    \caption{There are two modules in the IP verification protocol. The outer module is deployed outside the model API, responsible for generating triggers to obtain model copyright evidence and verifying the evidence. The inner module is added as a branch in the model API, responsible for checking whether the input is a trigger and generating the copyright evidence.}
    \label{fig:method}
\end{figure}

Figure~\ref{fig:method} shows the complete process of the branch watermarking protocol for the model. When an ownership dispute arises, the verifier uses the outer module to generate triggers, sends these triggers as query requests to the API, and verifies the returned responses to complete copyright certification. Specifically, when the API receives a query request, it forwards the input content to both the model module and the inner module (of the watermarking module), initially running in the service state. The model module performs the original inference process to obtain the output content. 

In the meantime, the inner module verifies if the current input is a trigger. If it is, the API transitions to the forensic state and proceeds to execute the evidence generator module. This module injects ownership information into the original output content of the model, obtaining ownership evidence. The API then returns this evidence as a response to the user.
If the input is not a trigger, the API continues in the service state, returning the model's original output as a response to the user. 
The triggers used in the forensic process are generated by the outer module. Importantly, the outer module is private or only open to a trusted third party.

Faced with the two situations of dishonest behavior in the threat model, the owner of Model A can naturally use the outer module to generate triggers, conduct investigations on suspicious remote application services, and verify the evidence. In the first case, sending triggers to the suspicious application that pretends to use Model A would yield no ownership evidence. In the second case, sending triggers to an application claiming to use Model B but with performance closely resembling Model A would yield the ownership evidence of Model A. 

\subsubsection{Defining Watermarking Protocol}
\label{section:watermarking-protocol}
Throughout this work, let $\lambda \in \mathbb{N}$ denote a security parameter, which is an implicit input for all the defined algorithms. A function $\epsilon$ is termed negligible if it approaches zero faster than any polynomial function. We represent a probabilistic polynomial-time algorithm as PPT. $API^W_M$ denotes the API of model $M$ after deploying watermarking module $W$, and when the existence of the inner watermarking module is unknown, it is represented as $API^*_M$. $W_{c,k}$ denotes the watermarking module that incorporates copyright information $c$ and key $k$ as internal parameters. The pair $(x, y)$ denotes a query-response pair for $API_M$, where $y = API_M(x)$, with $x \in X$ and $y \in Y$, where $X$ and $Y$ represent the input content space and response content space, respectively. $x_*$ represents a query $x$ that contains the message $*$, and $y_*$ represents a response $y$ that contains the message $*$, where $* \in \{tri, noi\}$, and $tri, noi$ denote meaningful triggers and meaningless noise, respectively.

\textbf{Definition 1. (Branch Backdoor-based Watermarking Protocol for Model)} A branch backdoor-based watermarking protocol for model is a set of possibly probabilistic algorithms, \textit{BranchWM = (KeyGen, TriggerGen, Branch, Deloy, Verify)}. 

\begin{itemize}
\item{\textit{KeyGen} is the key generation algorithm, taking an input of arbitrary content with a length of $\lambda$ and outputting a key $k$ used as the foundation for other algorithms.}
\item{\textit{TriggerGen} is the trigger generation algorithm, taking an input of arbitrary length query $x$ and key $k$ and producing a trigger version $x_{tri}$ for $x$. This corresponds to the trigger generator of the outer module.}
\item{\textit{Branch} is the watermarking algorithm, taking the $API^*_S$ of the suspicious model $S$ and trigger $x_{tri}$ as inputs and outputting the copyright evidence $(x_{tri}, y_*)$ for $API^*_S$, which is a key input for the \textit{Verify} algorithm. This corresponds to the inner module.}
\item{\textit{Deploy} is the watermarking deployment algorithm, taking the $API_M$, watermarking algorithm $W$, copyright information $c$, and key $k$ as inputs and generating the $API^{W_{c,k}}_M$ with encapsulated watermarking functionality. The watermarking-deployed $API^{BW}_M$ has both an original inference path $M$ and a backdoor watermarking path $BW$, which is as a parallel branch.}
\item{\textit{Verify} is the copyright verification algorithm, taking copyright information $c$, key $k$, and query-response pair $(x_{tri}, y_*)$ concerning $API^*_S$ as inputs and outputting whether the pair $(x_{tri}, y_*)$ is valid evidence. This corresponds to the evidence detector of the outer module.
If $y_*$ is $y_c$, the evidence is valid; otherwise, it is invalid.}
\end{itemize}

\textbf{Definition 2. (API State)} An API state represents the response status of an API when presented with different queries. 
The \textit{Service State} indicates that the watermarking function has not been activated, while the \textit{Forensic State} signifies that it has been activated. APIs deployed with watermarking algorithms have two response states—\textit{Service State} and \textit{Forensic State}—whereas original APIs always remain in the \textit{Service State}.

\textbf{Definition 3. (Branch Algorithm)} This is a supplementary definition of the \textit{Branch} algorithm in \textit{BranchWM}. It consists of two PPT sub-algorithms, simply denoted as \textit{BW = (Detect, Prove)}:

\begin{itemize}
\item{\textit{Detect} is the trigger detection algorithm, taking query $x_*$ and key $k$ as inputs and providing a detection result $s \in \{0,1\}$ regarding the presence of the trigger in the query.}
\item{\textit{Prove} is the copyright evidence generation algorithm, taking the original response content $y_{noi}$ of $API_M$, copyright information $c$, and key $k$ as input and outputting the response content $y_c$ with embedded $c$.}
\end{itemize}

For $API^{W_{c,k}}_M$, the output of the \textit{Detect} algorithm, $s(x_{noi}) = 0$, implies the non-activation of the watermarking function executed by the \textit{Prove} function. Thus, the API is in the \textit{Service State}, and $s(x_{tri}) = 1$ implies activation of the \textit{Prove} function, switching the API from the \textit{Service State} to the \textit{Forensic State}.

\textbf{Definition 4. (($\epsilon,\lambda$)-correct)} We call a \textit{BranchWM} ($\epsilon,\lambda$)-correct if a \textit{BranchWM} satisfy the following formulas:

\label{eq:1}
\begin{align}
\textbf{Verify}(k,(\textbf{API}^W_M(k,x_{tri}),x_{tri}))=1, \label{eq:1A}\\
\textbf{Verify}(k,(\textbf{API}_M(k,x_{tri}),x_{tri}))=0, \label{eq:1B}\\
\textbf{Verify}(k,(\textbf{API}^W_M(\neg k,x_{tri}),x_{tri}))=0, \label{eq:1c}
\end{align}

where $k$ is the output of $KeyGen(1^\lambda)$ and $\neg k$ is a random sequence whose length is equal to the length of $k$. $x$ is a normal sample and $x_{tri}$ is the output of $TriggerGen(x,k)$.

\subsubsection{Definition of Performance-lossless}
\label{section:performance-lossless}
We give the definition of performance-lossless for model watermarking, which intuitively means that the watermarking task does not affect the original task of the model. 

\textbf{Definition 5. (($\epsilon,t,q,\lambda$)-performance-lossless)}  We call a \textit{BranchWM} 
\textit{($\epsilon,t,q,\lambda$)-performance-lossless} if the deployed watermarking function only has a negligible impact on the performance of the protected model.  In other words, for any PPT algorithm $\mathcal{A}$, after $q$ queries on $API^{W_{\cdot,KeyGen(1^\lambda)}}_M$ and \textit{TriggerGen Oracle} within time $t$, $\mathcal{A}$'s distinguishing advantage between $API^{W_{\cdot,KeyGen(1^\lambda)}}_M$ and $API_M$ is negligible for the security parameter $\lambda$.

Since this paper does not consider adversaries with white-box capabilities, the performance-lossless property defaults to black-box performance-lossless.

\subsection{Provable Performance-lossless Watermarking from Secure Cryptographic Primitive}
\begin{figure}
    \centering
    \includegraphics[width=0.95\linewidth]{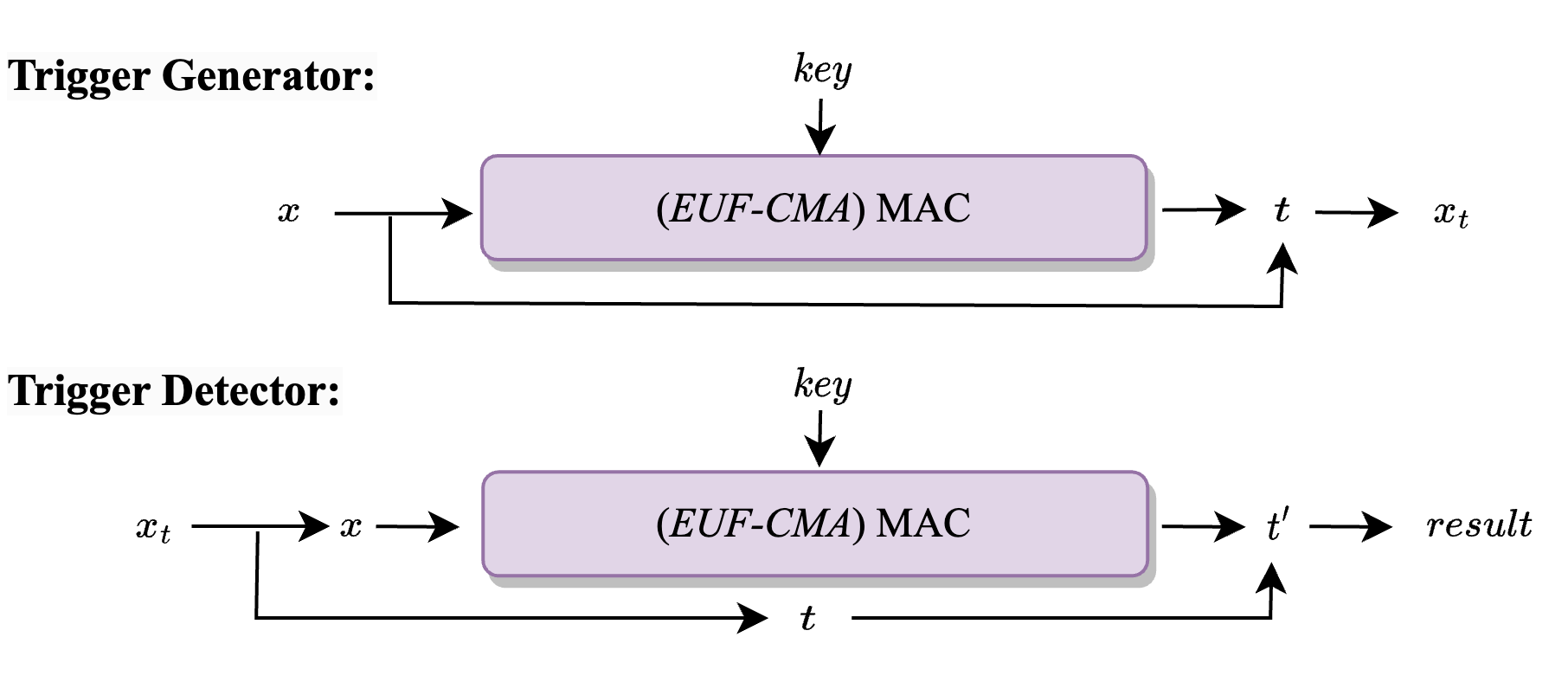}
    \caption{We use the MAC scheme which meets \textit{EUF-CMA} to construct the \textit{Trigger Generator} and \textit{Trigger Detector} sub-modules. Such a construction generates the MUFT that satisfies the \textit{DUL-CSA} property. }
    \label{fig:branch}
\end{figure}

Inspired by undetectable backdoors~\cite{goldwasser2022planting}, we utilize an MAC algorithm, which meets Existential Unforgeability under Chosen Message Attack (EUF-CMA), to construct the trigger generator and detector sub-modules. As shown in Figure~\ref{fig:branch}, the authentication code of the original sample is used as the triggering pattern, whose detection efficiency is better than that of the digital signature scheme based on asymmetric encryption technology. To better understand the relationship between the unforgeability of the MAC scheme and the performance-lossless property of the \textit{BranchWM}, we review some cryptographic primitives and propose the concept of Unaccessible Trigger (UT) based on them, whose pattern meets Distributional Unaccessibility under Chosen Sample Attack (DUL-CSA). Then, we use the MAC scheme to construct the forensic trigger which is named MAC-based unaccessible and fast trigger, i.e., MUFT. Finally, we prove that the proposed \textit{BranchWM} is performance-lossless by reduction.

\subsubsection{Cryptographic Primitive}
First, let us review some cryptography knowledge.

\textbf{Definition 6. (PseudoRandom Function, PRF)} For the function set $\mathcal{F}$, there is $\forall f \in \mathcal{F}, f : \{0, 1\}^\lambda \to \{0, 1\}^\lambda$, and the output distribution is $\mathcal{P}_\lambda$. For any probabilistic polynomial time (PPT) discriminator $\mathcal{A}$, if $\mathcal{A}$ adaptively submits multiple inputs $x_i$ to $\mathcal{F}$ and obtains the output $y_i$, the advantage of distinguishing situations I and II based on the output $y_i$ is negligible, then $\mathcal{F}$ can be said to be a PRF on $\mathcal{P}_\lambda$: 

\begin{flalign}
\label{eq:2}
\text{I: }y_i = f (x_i), f \gets \mathcal{F}, \\
\text{II: }y_i \gets \{0, 1\}^\lambda, \\
\textbf{Adv}^{PRF}_{\mathcal{A}}=\bigg|\mathop{Pr}\left[
\mathcal{A}(\text{I}) = 1 \right] - \mathop{Pr}\left[
\mathcal{A}(\text{II}) = 1 
\right]\bigg|.
\end{flalign}

\textbf{Definition 7. (Message Authentication Code, MAC)} MAC can be expressed as a tuple consisting of three polynomial time algorithms \textit{(Key, Mac, Veri)}, defined as follows:

\begin{itemize}
\item{\textit{KeyGen} is the key generation algorithm, uniformly outputting a key $k$ of length $\lambda$ according to the selected security parameter $\lambda \gets \textit{Key}(1^\lambda)$;}
\item{\textit{Mac} is the message encoding algorithm, taking $k \in \{0, 1\}^\lambda$ and $M \in \{0, 1\}^L$ as inputs and outputting $\textit{tag} \in \{0, 1\}^l$, where tag is the message authentication code corresponding to $M$, that is, $\textit{tag} \gets \textit{Mac}_k(M)$;}
\item{\textit{Veri} is the message verification algorithm, taking $k \in \{0, 1\}^\lambda$, $\textit{tag} \in \{0, 1\}^l$, and $M \in \{0, 1\}^L$ as inputs and outputting one bit $b \in \{0, 1\}$; $\forall k$ and $M$, there is $\textit{Veri}_k(\textit{Mac}_k(M), M)) = 1$.}
\end{itemize}

If the MAC scheme satisfies the PRF security in Definition 6, it satisfies \textit{EUF-CMA}, which is defined as follows.

\textbf{Definition 8. (Existential Unforgeability under Chosen Message Attack, EUF-CMA)} For an MAC algorithm \textit{MAC} = (\textit{Key}, \textit{Mac}, \textit{Veri}), any PPT adversary $\mathcal{A}$ can select message $M_i \in \{0, 1\}^l$ and query the \textit{MAC} for $\textit{tag}_i = \textit{Mac}_k(M_i)$ within time $t$. After $q$ queries, the \textit{MAC} is said to satisfy \textit{EUF-CMA} security, if the advantage that $\mathcal{A}$ outputs $(M',\textit{tag}')$ where $M' \notin \{ M_1, \cdots, M_q\}$, $\textit{Veri}_k(\textit{tag}', M') = 1$, is negligible:

\label{eq:3}
\begin{align}
\mathop{Max}_{A \in \mathcal{A}(t,q)}\{\textbf{Adv}^\textit{EUF-CMA}_\mathcal{A} \}\leq \epsilon(\lambda).
\end{align}

Then, \textit{MAC} can be called $(\epsilon, t, q, \lambda)$ - secure \textit{MAC}.

\subsubsection{Unaccessible Trigger}
We construct the \textit{Trigger Generator} and \textit{Trigger Detector} sub-modules in the \textit{BranchWM} based on the MAC scheme. If the MAC scheme satisfies $(\epsilon, t, q, \lambda)$ - security in Definition 8, the trigger pattern of \textit{BranchWM} satisfies the Distributional Unaccessibility under Chosen Sample Attack (DUL-CSA), which is defined as follows.

\textbf{Definition 9. (Distributional Unaccessibility under Chosen Sample Attack, DUL-CSA)} 
For a \textit{BranchWM} = (\textit{KeyGen}, \textit{TriggerGen}, \textit{Branch}, \textit{Deploy}, \textit{Verify}), any PPT adversary $\mathcal{A}$ in the ``real'' world can select sample $x_i \in \{0, 1\}^l$ and query $\mathcal{O}^{TG}$ (i.e., the trigger generator of the protocol's outer module)  for $x_{i,trigger} \gets \textit{TriggerGen}_k(x_i)$ within time $t$. After $q$ queries, \textit{BranchWM} is said to satisfy $(\epsilon, t, q, \lambda)$\textit{-DUL-CSA} property, if the advantage, that $\mathcal{A}$ output $(x',x_{tri}')$ where $x' \notin \{ x_1, \cdots, x_q\}$, $\textit{Detect}_k(x_{tri}',) = 1$, is negligible:

\begin{flalign}
\label{eq:2}
\text{I: }KeyGen(1^\lambda) = k, \mathcal{A}^{\mathcal{O}^{\textit{TG}_k}}(\mathcal{O}^{API}) = x',x_{tri}', \\
\text{II: }KeyGen(1^\lambda) = k, 
\mathcal{A}(\mathcal{O}^{API}) = x',x_{tri}', \\
Pr(\text{real}) = Pr(Detect_k(x_{tri}') = 1 |\text{I}), \\
Pr(\text{ideal}) = Pr(Detect_k(x_{tri}') = 1 |\text{II}), \\
\textbf{Adv}^{UT}_{\mathcal{A}}=\bigg|Pr(\text{real}) - Pr(\text{ideal})\bigg|, \\
\mathop{Max}_{A \in \mathcal{A}(t,q)}\{\textbf{Adv}^\textit{UT}_\mathcal{A} \}\leq \epsilon(\lambda).
\end{flalign}

where ``real'' means that $\mathcal{A}$ has access to the model API (i.e., $\mathcal{O}^{API}$) as well as oracle access to the trigger generator of the protocol's outer module (i.e., $\mathcal{O}^{Act}$) to which she or he can make queries; ``ideal'' means that $\mathcal{A}$ can only access $\mathcal{O}^{API}$.

Then, the trigger pattern of the \textit{B ran ch WM} can be called a $(\epsilon, t, q, \lambda)$-secure trigger pattern. We refer to such triggers as unaccessible triggers. Because they are constructed from the MAC scheme, we also call them MAC-based unaccessible and fast triggers, i.e., MUFTs, where the ``fast'' is discussed in Section~\ref{section:runtime}.

\subsubsection{Proof}
The idea of multi-branch structure separates the model's original task and forensic task. Unless a branch is activated by a trigger, the model will maintain its initial inference results. This means that the ideal model fidelity is determined by the reliability of the trigger mechanism; that is, the trigger detector of \textit{BranchWM} should have a negligible false positive rate (i.e., robustness) and a negligible false negative rate (i.e., completeness). The significance of an unaccessible trigger pattern is that even if an attacker deliberately constructs a query, they cannot activate the trigger detector without the correct key used in trigger generation.
Therefore, if the trigger pattern of \textit{BranchWM} satisfies ($\epsilon,t,q,\lambda$)\textit{-DUL-CSA} in Definition 9, the trigger detector submodule satisfies completeness and soundness. Then, a \textit{BranchWM} constructed from a MAC scheme that meets ($\epsilon,t,q,\lambda$)\textit{-EUF-CMA} is a ($\epsilon,t,q,\lambda$)-performance-lossless \textit{BranchWM}, which is proven as follows:

\textbf{Completeness.} This property is satisfied based on a proof of correctness of the MAC scheme. Tags generated by a \textit{Mac} function under the same symmetric key and the same message can always be the same. Therefore, as shown in Fig~\ref{fig:branch}, triggers generated by the \textit{Trigger Generator} can always be detected by the \textit{Trigger Detector} under the symmetric key without considering the possibility of an attacker distorting the triggers.

\textbf{Soundness.} This property is reduced to the security of the MAC scheme. If there exists a PPT algorithm $\mathcal{A}$ that can generate a trigger query $x'_{{trigger}}$ without knowledge of the symmetric key $k$ and successfully obtain $\textit{Detect}_k(x_{tri}',) = 1$, then $\mathcal{A}$ can be used to forge a valid $\textit{tag}$ on $x'$ about $k$. This contradicts the \textit{DUL-CSA} security of the trigger pattern, thereby contradicting the \textit{EUF-CMA} security of MAC. Hence, without the knowledge of $k$, the probability that trigger queries can be successfully forged is negligible. Not to mention the possibility of triggering a forensic state with unintentional queries.

\subsection{Comparison of Secure Cryptography Primitives}
\subsubsection{Secure Hash Function}
A hash function $h$ is an algorithm that maps arbitrary length data to fixed length data.\footnote{This is an informal definition of an individual hash function, not a hash function family. The hash function used in practice can be regarded as a hash function with determined parameters in the hash function family.}
For any input data, a hash function can generate a unique hash value. Generally speaking, a secure hash function has the following properties:

\begin{itemize}
 \item{\textit{Preimage resistance:} For any given hash output $y$, it is computationally infeasible to find a corresponding message input $x$ such that $h(x)=y$.}
 \item{\textit{Second preimage resistance:} For any given input $x$, it is computationally infeasible to find an input $x'$ that is different from x such that $h(x)=h(x')$.}
 \item{\textit{Collision resistance:} It is computationally infeasible to find two different inputs $x$ and $x'$ such that $h(x)=h(x')$.}
\end{itemize}

\subsubsection{EUF-CMA Signature}
A digital signature is a cryptographic technology used to verify and ensure the origin and integrity of digital information. It is based on asymmetric encryption and usually includes three algorithms: key generation, signing, and verification. To ensure security, existential unforgeability is an important security property of a digital signature scheme:

\begin{itemize}
 \item{\textit{Existential unforgeability:} Without holding the private key, an adversary with polynomial computing power, after querying a certain number of (message, signature) pairs, is still unable to construct a valid signature for a message $m$ that has not been previously queried for any corresponding signature.}
\end{itemize}

\begin{table}[ht]
\renewcommand{\arraystretch}{1.5}
\centering
\setlength{\abovecaptionskip}{1pt}
\setlength{\belowcaptionskip}{1pt}
\setlength{\tabcolsep}{3pt}
\caption{The runtime cost of the verification time between different cryptographic primitives. We chose one hash function (i.e., SHA512), one MAC scheme (i.e., HMAC\_SHA512), and two digital signature schemes (i.e., ECDSA and FALCON) according to the recommendations of NIST~\cite{nist_hash,nist_hmac,nist_sig,PQC}. We use SHA512 as the baseline to record the ratios of HMAC, ECDSA and FALCON under the 1024-bit, 163-bit and 10248-bit key lengths, respectively. 
FALCON\_512 is its basic security version. }
\begin{tabular}{cccccc}
\toprule
Method & Key Size (bit) & Output Size (bit) & Time Rate \\ 
\hline
SHA512 & $ / $ & $ 512 $ & $ 1 (0.0114ms) $ \\
HMAC\_SHA512 & $ 1024 $ & $ 512 $ & $ \bm{1.2020_{\pm 0.0784}} $ \\
ECDSA\_SHA512 & $ 163 $ & $ 326 $ & $ 64.6781_{\pm 2.1548} $ \\ 
FALCON\_512 & $ 10248 $ & $ 5520 $ & $ 5.2550_{\pm 0.1899} $ \\ 
\bottomrule

\end{tabular}
\label{tab:crypt-time}
\end{table}

To improve security, the signing algorithm is usually a probabilistic algorithm that introduces randomness into the signature generation process so that different signatures can be obtained when the same message is signed multiple times. The above security is a kind of weak existential unforgeability\footnote{The existence unforgeability of MAC defined in this paper is weak existence unforgeability.}, and the adversary's assumption is that it has never observed any signature of the message $m$.
Accordingly, strong existential unforgeability means that after querying the signature of the message $m$, the attacker still cannot generate a valid new signature for it.

\subsubsection{Hash vs MAC vs Signature}
\label{section:runtime}
\textbf{Runtime Cost.} Goldwasser et al.~\cite{goldwasser2022planting} used a strong unforgeable digital signature scheme to construct the trigger mechanism of the backdoor and further combined it with the Cook-Levin theorem to achieve the non-replicability (i.e., unforgeability) of the trigger in a white-box model situation. However, as shown in Table~\ref{tab:crypt-time}, digital signature schemes based on asymmetric cryptography will introduce heavier operating overhead in the service state of the model API than hash functions and MACs.\footnote{FALCON is a lattice-based post-quantum signature algorithm. Although it achieves high enough security with a short signature length and fast speed in the cryptograhic field, this signature size is still unacceptable as a trigger pattern of the language model watermarking.}
In addition, in the case of the white-box model, the model will face an easier but more severe attack—the adversary can identify and remove the backdoor or watermark based on structural anomalies. The benefits of white-box unforgeability are not sufficient to extend the capability of an adversary from the black-box to the white-box.

\textbf{Trigger Pattern Space.} Strong unforgeable digital signatures can generate multiple trigger patterns for the same original sample $x$, making the trigger pattern space larger. However, a large pattern space is not necessary. A large sample space can guarantee a large trigger space. It is feasible to generate a unique trigger pattern for $x$ based on a hash function or MAC and record $x$ to achieve a one-time trigger. 

\textbf{Security.} Hash functions are the most efficient of the three cryptographic primitives. If a secure hash function is used to construct the trigger mechanism of the branch backdoor-based model watermarking, the running overhead introduced by the deployed watermark is small when the model API is in the service state. However, non-key-driven hash functions make it hard to protect the model without keeping the hash function secret.

Therefore, we chose an efficient and secure MAC to construct a performance-lossless branch watermarking scheme for model.

\section{Performance-lossless and Secure Watermarking for Language Model}
In this section, we provide a concrete example of the branch backdoor-based watermarking protocol for model using a language model as a case study. Then, we further analyze the possible attacks when the threat model faces the watermarking protocol during the forensic phase and propose a more secure instantiation strategy. 

\begin{algorithm}
\caption{TriggerGen Function}
\label{alg:alg1}
\begin{algorithmic}
\Require Prompt $x$, MAC Key $k$
\Ensure  Triggered Prompt $x_{tri}$

\State $\sigma \gets \text{Mac}(k,x)$ \Comment{Mac Function of MAC}
\State $m \gets \sigma$, $n \gets 0$, $t\gets []$, $v \gets | \mathcal{V} |$
\While{$m>0$} \Comment{\textit{Embed} Function}
\State $n \gets m\mod{v}$
\State $m \gets m//{v}$
\State $t \gets t+[n]$ \Comment{Add to list}
\EndWhile
\State $t \gets \text{tokenizer.decode}(t)$
\State $x_{tri} \gets x || t$

\State \textbf{Output}  $x_{tri}$
\end{algorithmic}
\label{alg1}
\end{algorithm}

\begin{algorithm}
\caption{Detect Function}
\renewcommand{\algorithmicrequire}{\textbf{Input:}}
\renewcommand{\algorithmicensure}{\textbf{Output:}}
\label{alg:alg2}
\begin{algorithmic}
\Require Prompt $x_*$, MAC Key $k$
\Ensure  Trigger Detection Result $r$

\State $v \gets | \mathcal{V} |$, $\sigma \gets 0$
\State Parse $x_*$ as ${x^0,x^1}$
\State $t \gets\text{tokenizer.encode}(x^1)$
\For{$i\in \{0,1,...,|t|-1\}$} \Comment{Extract Function}
\State $\sigma \gets \sigma + v^i*t[i]$
\EndFor

\State $r \gets \text{Veri}(k, x^0,\sigma)$ \Comment{Veri Function of MAC}

\State \textbf{Output}  $r$
\end{algorithmic}
\label{alg2}
\end{algorithm}

\begin{algorithm}
\renewcommand{\algorithmicrequire}{\textbf{Input:}}
\renewcommand{\algorithmicensure}{\textbf{Output:}}
\caption{Prove Function}
\label{alg:alg3}
\begin{algorithmic}
\Require Trigger Detection Result $r$, Original Response $y$
\Ensure  Response $y'$

\State $y' \gets \text{``I am model B from owner A!''} $
\If {$r=0$}
\State $y' \gets y$
\EndIf
\State \textbf{Output}  $y'$
\end{algorithmic}
\label{alg3}
\end{algorithm}

\subsection{Performance-lossless Watermarking for Language Generation Model}
In this section, we take the language model as an example to describe a \textit{performance-lossless BranchWM}  based on the unforgeable MAC scheme. 

\subsubsection{Language Model}
The language model, denoted as $LM$, is a deterministic algorithm that iteratively predicts probability distributions on a specific set of tokens, $T$. It takes an input consisting of a prompt (i.e., query) sequence, $x$, and the token sequence obtained from previous iterations, $t = (t_1, ..., t_{i-1})$. This input can be referred to as the history, denoted as $h = (x, t)$. The output of the language model is a probability distribution on the set of tokens $T$, often represented as $prob_i = LM(h)$.

In practical usage, when the initial input is given as $t = ()$, the output can be more generally represented as $logit_i$, which has not yet been processed into $prob_i$ yet. The probability distribution $prob_i$ is typically obtained by applying the Softmax function to $logit_i$, which normalizes the logits into probabilities.

\subsubsection{Watermarking Instance}
\label{section:watermarking-instance}
Algorithm~\ref{alg:alg1} defines the \textit{TriggerGen} function, which is responsible for generating triggers using a digital signature scheme. It combines the message encoding function \textit{Mac} and an embedding function \textit{Embed}. The input is an arbitrary-length prompt $x$ and the signing key $sk$. It first applies the \textit{Mac} function to obtain a signature $\sigma$ of the prompt $x$ in token form using the signing key $sk$. Then, it uses the \textit{Embed} function to convert the signature $\sigma$ (i.e., trigger) into token form, which is concatenated to the end of the original prompt $x$. The final output is a trigger version $x_{tri}$ for $x$.

Unlike image pixels, both text and tokens lack the ideal redundancy to carry triggers. Here, we implement a simple \textit{Embed} function that converts the trigger directly into the same form as the signed message and concatenates them. This approach is feasible without considering the possibility of an attacker detecting and intercepting anomalous queries.

The \textit{Verify} algorithm is used to confirm whether the response of the suspicious API to the trigger query is the statement set in the Prove function.

Algorithms~\ref{alg:alg2} and~\ref{alg:alg3} define the \textit{Branch} function. It includes the trigger detection algorithm \textit{Detect} and the evidence generation algorithm \textit{Prove}. The \textit{Detect} function uses the verify algorithm \textit{Veri} corresponding to the message encoding function \textit{Mac} used in the \textit{TriggerGen} function. The symmetric key $k$ is the implicit input of the \textit{Veri} function. It takes a prompt $x_*$ as input, converts $x_*$ into a binary sequence, and extracts a fixed-length sequence from the end as a potential trigger (i.e., signature), $\sigma$. The remaining part of $x_*$ is $x$. Then, the \textit{Veri} function is used to check whether $\sigma$ is a valid message tag of $x$ under the symmetric key $k$. Determine whether to activate the \textit{Prove} function based on the trigger detection results. Here we implement a simple \textit{Prove} function that replaces the model's inference results directly with statements such as \textit{``I am model B from owner A!''} The content format can vary, including token sequences, sharp probability distributions, or logical values, depending on the API's output format. 

\begin{figure}
    \centering
    \includegraphics[width=0.95\linewidth]{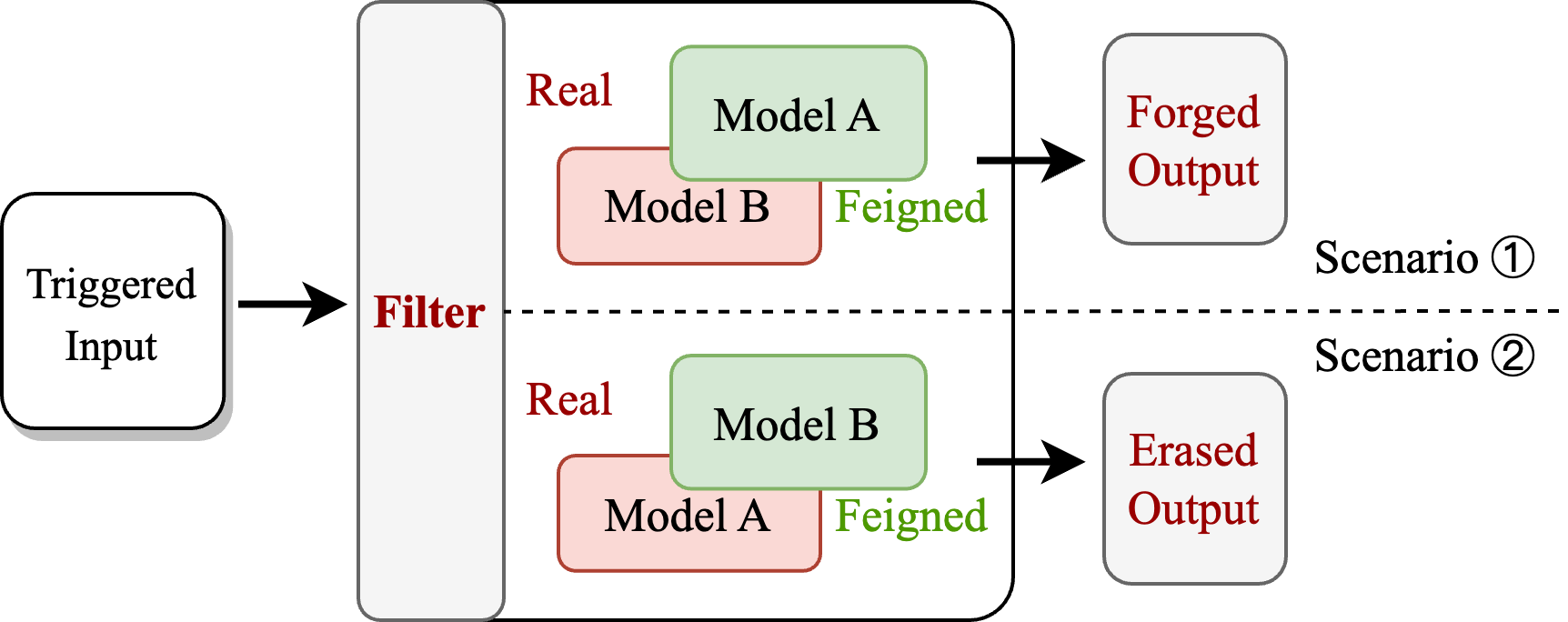}
    \caption{Interference attacks in two scenarios: (1) The attacker uses a low-cost model B to feign a high-cost model A and tries to filter the input or forge the output; (2) The attacker lies about having self-developed high-cost model B which is actually model A and tries to filter the input or erase the watermark signal in the output. }
    \label{fig:attack}
\end{figure}

\subsection{Interference Attacks}
\label{section:interference-attacks}
In practical use, the branch backdoor-based watermarking protocol for model may face targeted interference at tacks against the forensic (i.e., evidence collection) process from attackers when dealing with two types of dishonest behaviors in the threat model. Figure~\ref{fig:attack} shows the following two attack scenarios:

\textbf{Scenario 1: A low-cost model B pretends to be a high-cost model A.} Attackers attempt to detect triggers at the input side to produce appropriate responses at the output side. A straightforward way is to reject triggers as illegal inputs. However, the number of triggers is enormous and rejecting a single trigger is reasonable, but the act of rejecting all triggers may itself be suspicious. Therefore, attackers need to take potential trigger requests seriously. When attackers can access high-cost model A, the trigger is detected and then sent to A, returning the specific copyright evidence, thus successfully deceiving at a critical moment in the forensic process. When attackers cannot access A, they detect triggers and attempt to forge evidence.

\textbf{Scenario 2: Lie about having self-developed high-cost model B, but it is actually model A.} Attackers attempt to detect triggers at the input side to produce appropriate responses at the output end. When attackers have a low-cost alternative model C, they send the detected triggers to it to return specific outputs, indicating that the triggers have no effect on B, thus demonstrating that there is no copyright dispute. When attackers do not have model C available, they send the detected triggers to A to obtain specific copyright evidence and perform an evidence erasure attack to achieve the same purpose. Additionally, attackers can also detect copyright evidence at the output side and return the same alternative or erased output.

\subsection{Analysis for Interference Attacks}
We summarize the principles for achieving resistance against interference attacks as follows:
\begin{enumerate}
    \item Triggers and copyright evidence are concealed and diverse, making it difficult for attackers to detect and intercept;
    \item Copyright evidence is hard to forge;
    \item Triggers and copyright evidence are robust, enabling them to withstand a certain degree of erasure by attackers.
\end{enumerate}
Notably, an attacker can distort the trigger to disable it, thereby bypassing the evidence collection process. However, in the text generation task of language models, tampering with the input brings more uncertainty than erasing the known output, which may seriously affect the attacker's original service quality. Therefore, Principle (3) can be relaxed to only emphasize the robustness of copyright evidence.

Obviously, the current output of the \textit{TriggerGen} function, algorithm~\ref{alg:alg1}, lacks stealth. The \textit{Embed} function directly converts the signature sequence into a token sequence with no natural semantics, resulting in garbled characters after being appended to the original prompt. It can be easily detected using simple perplexity detection. The inference output content replaced by the \textit{Prove} function, algorithm~\ref{alg:alg3}, is also easy to detect and forged by attackers, lacking stealth and robustness.

\subsection{Improving Security of Watermarking}

In this section, we improve upon the previously proposed simple structure to resist potential interference attacks during evidence collection.

We show how to improve the concealment and robustness of the \textit{TriggerGen}, \textit{Detect}, and \textit{Prove} functions constructed in section~\ref{section:watermarking-instance} by integrating existing no-box watermarking schemes, including a robust multi-bit watermarking method~\cite{yoo2023color} and a simple multi-bit watermarking method modified from Kirchenbauer et al.'s work~\cite{kirchenbauer2023watermark}.

\begin{algorithm}[H]
\renewcommand{\algorithmicrequire}{\textbf{Input:}}
\renewcommand{\algorithmicensure}{\textbf{Output:}}
\caption{Concealed TriggerGen: Embedding Function}
\label{alg:alg4}
\begin{algorithmic}
\Require Prompt $x$, Signature $\sigma$, Model $\mathcal{M}$, Vocabulary $\mathcal{V}$
\Ensure  Triggered Prompt $x_{tri}$

\State Parse $\sigma$ as $\{\sigma_0,\sigma_1,...,\sigma_{|\sigma|-1}\}$
\State $history \gets \text{tokenizer.encode(x)}$
\State $logit \gets \mathcal{M}(history)$
\State $prefix \gets \text{Sample}(\text{Softmax}(logit),\mathcal{V})$
\State $history \gets history + [prefix]$

\For{$i\in \{0,1,...,|\sigma|-1\}$}
\State $seed \gets \text{Hash}(prefix)$
\State $\mathcal{V}^0, \mathcal{V}^1 \gets \text{Permutation}(seed, \mathcal{V})$ 
\Comment{Permutation and Divide the Vocabulary}
\State $logit \gets \mathcal{M}(history)$
\State $prefix \gets \text{Sample}(\text{Softmax}(logit),\mathcal{V}^{\sigma_i})$
\State $history \gets history+[prefix]$ \Comment{Add to list}

\EndFor
\State $x_{tri} \gets \text{tokenizer.decode}(history)$

\State \textbf{Output}  $x_{tri}$
\end{algorithmic}
\label{alg4}
\end{algorithm}

Algorithm~\ref{alg:alg4} defines the concealed version of the \textit{Embed} function in the \textit{TriggerGen} function. A concealed \textit{Embed} function is implemented using the watermarked text generation algorithm of the semantic-preserving no-box watermarking scheme modified from Kirchenbauer et al.'s work~\cite{kirchenbauer2023watermark}. It takes an arbitrary-length prompt $x$ and the signature $\sigma$ of $x$ as input, with the embedding key $ek_{in}$ as an implicit input. The binary sequence of the signature $\sigma$ is embedded into a newly generated text  that has $x$ as context (i.e., initial history). Then, the newly generated text is appended to $x$ to produce $x_{tri}$.
Table~\ref{tab:text-pre} displays the simple triggers from algorithm~\ref{alg:alg1} and the concealed triggers generated from algorithm~\ref{alg:alg4}. It shows that the latter maintains semantic readability and has good concealment, while the former is almost garbled.

\begin{algorithm}[H]
\renewcommand{\algorithmicrequire}{\textbf{Input:}}
\renewcommand{\algorithmicensure}{\textbf{Output:}}
\caption{Concealed Detect: Extract Function}
\label{alg:alg5}
\begin{algorithmic}
\Require Prompt $x_*$, Vocabulary $\mathcal{V}$
\Ensure  Extracted Trigger $\sigma$

\State $\sigma \gets []$
\State Parse $x_*$ as ${x^0,x^1}$
\State $prefix \gets \text{tokenizer.encode}(x^0)[-1]$
\State $t \gets\text{tokenizer.encode}(x^1)$
\For{$i\in \{0,1,...,|t|-1\}$} 
\State $seed \gets \text{Hash}(prefix)$
\State $\mathcal{V}^0, \mathcal{V}^1 \gets \text{Permutation}(seed, \mathcal{V})$ 
\Comment{Permutation and Divide the Vocabulary}
\State $b \gets \text{GetIndex}(\mathcal{V}^0,\mathcal{V}^1,t[i])$ \Comment{Get the vocabulary index that the current token belongs to}
\State $\sigma \gets \sigma + [b]$ \Comment{Add to list}
\State $prefix \gets t[i]$
\EndFor

\State \textbf{Output}  $\sigma$
\end{algorithmic}
\label{alg5}
\end{algorithm}

\begin{algorithm}[H]
\renewcommand{\algorithmicrequire}{\textbf{Input:}}
\renewcommand{\algorithmicensure}{\textbf{Output:}}
\caption{Concealed Prove: Embedding Function}
\label{alg:alg6}
\begin{algorithmic}
\Require Trigger Detection Result $r$, Original Logit $y$, Signature $\sigma$, Prefix $prefix$, Copyright Information $c$, Embedding State $(\delta,j)$, MAC Key $ek$, Vocabulary $\mathcal{V}$
\State \Comment{$\delta$ is the embedding strength,$j$ is the embedding length.}
\Ensure  Logit $y'$

\State $y' \gets y$
\State $\sigma' \gets \text{Mac}(k, \sigma)$
\State $\sigma'_1, \sigma'_2 \gets \sigma'$ \Comment{Divide $\sigma'$ into two parts from the middle}
\If {$r=1$}
\State $seed \gets \text{Hash1}(\sigma'_1||prefix)$
\State $\mathcal{V}^0, \mathcal{V}^1,...,\mathcal{V}^{Ceil(|c|/j)} \gets \text{Permutation}(seed, \mathcal{V})$ \Comment{Permutation and Divide the Vocabulary}
\State $start \gets \text{Hash2}(\sigma'_2||prefix) \mod j$
\State $pos \gets c[start:start+j]$
\State $y'[\mathcal{V}^{pos}] \gets y'[\mathcal{V}^{pos}] + \delta$
\EndIf

\State \textbf{Output}  $y'$
\end{algorithmic}
\label{alg6}
\end{algorithm}

Algorithm~\ref{alg:alg5} redefines the \textit{Extract} function in the \textit{Detect} function to adapt to the improved concealed \textit{Embed} function in the \textit{TriggerGen} function. The extraction operation is implemented using the watermarked text extraction algorithm of the semantic-preserving no-box watermarking scheme modified from Kirchenbauer et al.'s work~\cite{kirchenbauer2023watermark}. It takes the prompt $x_*$ as input, with the embedding key $ek$ as an implicit input. The prompt $x_*$ is hashed token-by-token to extract the potential signature $\sigma$ and the original prompt $x$.

Algorithm~\ref{alg:alg6} defines the concealed and robust version of the \textit{Prove} function. A concealed and robust \textit{Prove} function is implemented using the logit modification algorithm of the semantic-preserving and robust no-box watermarking scheme~\cite{yoo2023color}. The pseudocode of the algorithm shows the modification of the original model inference result $y_i$ by the \textit{Prove} function during the single-step token generation process. At each step of token generation, the copyright information $c$ is partially embedded into the modified reasoning result $y_i'$. This algorithm is called repeatedly until the complete response content is generated. The complete response process takes the copyright information $c$ and the original model's inference result $y$ as input, with the embedding key $ek_{out}$ as an implicit input. Then, $c$ is embedded into $y$ to obtain a modified inference result $y_c$, which is then output as the final inference result.

The complete \textit{Verify} algorithm includes copyright information extraction and verification. 
The extraction function can be implemented using the extraction algorithm~\cite{yoo2023color} corresponding to Algorithm~\ref{alg:alg6}.
During the forensics process, infringement is confirmed when the \textit{Verify} algorithm successfully extracts and verifies the suspect API's response to the trigger.

\subsubsection{Security Analysis}
\textbf{The triggers and copyright evidence are concealed and diverse.} The semantic-preserving watermarking scheme ensures that the triggers and copyright evidence are concealed. The prompt space provides a variety of options for the original prompt $x$. The signing function can be a probabilistic algorithm that provides multiple signatures for a message. When $x$ has high entropy, it can generate diverse triggers with consistent context but varied expressions. Diverse triggers can provide various copyright evidence. Therefore, the triggers and copyright evidence are diverse.

\textbf{Copyright evidence is difficult to forge.} Without knowledge of the embedding key, it is challenging for an attacker to create evidence from which valid copyright information can be extracted. The evidence needs to have consistent context and natural semantics with the trigger.

\textbf{Copyright evidence is robust.} Taking advantage of the inherent robustness of the concealed watermarking scheme, a certain degree of insertions, deletions, and replacements can be tolerated. Notably, if an attacker attempts to make large-scale and high-quality modifications to copyright evidence, the costs will outweigh the benefits.

\begin{table*}[h]
\centering
\caption{Examples of simple triggers and concealed triggers (truncated for display). Take the single sentence in the story reasoning dataset ROCStories~\cite{rocstories} as pormpt, HMAC\_SHA512 generates the corresponding trigger pattern, and uses the tokenizer and model of OPT-1.3B~\cite{zhang2022opt} to generate triggers. }
\label{tab:text-pre}
\setlength{\tabcolsep}{2mm}
\renewcommand{\arraystretch}{1.5}
\noindent
\begin{tabularx}{\textwidth}{|p{0.15\textwidth}|p{0.25\textwidth}|X|}
\hhline{|-|-|-|}
\multicolumn{1}{|c|}{\textbf{Prompt}} & \multicolumn{1}{c|}{\textbf{Simple Trigger}} & \multicolumn{1}{c|}{\textbf{Concealed Trigger}} \\
\hline
Ben\textbackslash's Boy Scout Troop worked for weeks on a float. &  BuddyWC cohorts indefinite settle specialschurch visHallBreakuta ost fragmentationfur preachrows Previousagonsaum1100 KILLgru Fren Essentiallyateurs adeptroadilers LOVE dealershipXMuls Sch & Then it was destroyed by an avalanche. It\textbackslash's a tragedy.\textbackslash n"It\textbackslash's heartbreaking to think that our troop had so many great memories, but then they were all destroyed by an avalanche, and we don\textbackslash't have anything left of them. All the memories, all the work that our boys put in, it was taken from them by an avalanche, and that\textbackslash's terrible. And that\textbackslash's why our troop has been chosen to help the families of the deceased and to help the people who... \\
\hline
Anna was filling her bird feeders. & any committees foo labour agreeable Rhodcli bisexual Armory--ruptedUSH corporepadorter JacquesDenver width professional turnbenefit consolidation Kurinarily coronboxes89 carb Mk FistClick Smash 100 & A little bird flew in, looked at her for a few, then turned and flew out.\textbackslash nThat's when she realized there was a bird in the nest next door.\textbackslash nShe didn't know what it looked like, so she called to it.\textbackslash nThe little bird hopped from tree to tree and finally settled on a tree branch next to where she was feeding the birds, just as the sun was shining through the window on the first day of winter. Anna looked down and saw a blackbird sitting on a branch,... \\
\hline
Oscar decided that he no longer wanted to be out of shape. He had to change his body. & Senegal1972 Guests stalls brown sudden Macronalysedjandro Rip Manual Anthrop Cooperative $\sim$/. discrepanciesamon Null Stim \#\#\# railwaysicy recountottage346 onsetangingOil Ber SOC Where limited deteriorating Netflix & He had to train hard to get back to what he was before.\textbackslash nI\textbackslash'm so excited to see the results.\textbackslash nHe\textbackslash's been working hard for the past year to get back on track. We are happy to hear that he feels better about himself. He wants to prove to everyone that he can be the best fighter in MMA.\textbackslash n"It\textbackslash's a new year and everything, and I think it\textbackslash's important to make changes, especially in your body because you\textbackslash're going to be fighting every three months... \\
\hhline{|-|-|-|}
\end{tabularx}
\end{table*}

\section{Discussion and Limitations}

\subsection{Discussion}
\subsubsection{Replay Attacks}
The construction in this paper is based on message authentication technology in cryptography, which naturally raises the question of \textit{replay attacks}, a threat faced by MAC. In this paper, due to the vulnerability of MAC to replay attacks, attackers can execute trigger replay attacks. However, this type of attack is of limited practical significance. 

We envision an alternative replay attack called the \textit{``copyright evidence replay attack''} for the proposed scheme. While attackers cannot forge valid copyright evidence for a new trigger, they can search for semantically similar triggers they have encountered before and reuse their copyright evidence as a response to the new trigger. This attack highlights a weakness of the BranchWM; that is, the correlation between the trigger input and the evidence output is relatively weak, and semantic relevance alone is insufficient.

To solve this problem, we draw inspiration from MAC technology solutions to incorporate timestamps into copyright evidence to maintain temporal correlation beyond semantic correlation. Another idea is to hash the carried tag $\sigma$ and use it as the embedded key $ek$ for the \textit{Prove} function, as shown in Algorithm~\ref{alg:alg6}. This strong connection between triggered prompt and copyright evidence increases resistance to forgery and replay attacks.

\subsubsection{Generalization}
This paper introduces a general watermarking protocol framework for model in black-box scenarios. We give a branch indicator based on the MAC scheme. In fact, the MAC scheme can be replaced by any cryptographic primitive with unforgeability security to adapt to specific scenarios. In other words, the underlying cryptographic primitives are flexible and extensible. In addition, while the watermarking example is constructed for pre-trained language models in text generation tasks, the framework is applicable to various media formats, such as images and audio, and is not limited by network structures or tasks. As long as the actual application requirements are met, appropriate trigger embedding and extraction functions and representations of copyright evidence can be designed.

\textbf{Image Classification Task.} Different from the text generation task exemplified in Section 4, images are more redundant than text, and their embeddable space can meet the length requirements of secure cryptographic primitives. The following is a simple example of performance-lossless branch watermarking for an image classification model: within the TriggerGen algorithm, the specific pixel plane of the original image is input into the Mac algorithm to generate a trigger pattern, and then a simple Least Significant Bit ( LSB ) replacement method is used to embed the pattern into the image to obtain a trigger image; within the Detect algorithm, the tag of the specific pixel plane is recalculated and then compared with the LSB plane. If they are the same, the image is a trigger image; within the Prove algorithm, the model output (i.e., prediction label) is replaced with the last bit value of the trigger pattern; the Verify algorithm detects whether the label of the trigger image predicted by the suspicious API is equal to the last bit value it carries.

\subsection{Limitations}
\subsubsection{Model Extraction Attacks}
The \textit{performance-lossless} property defined in this paper for model watermarking contradicts the defense against model extraction attacks in black-box scenarios. In a model extraction attack, attackers aim to obtain a training dataset by querying the victim model to train an alternative model capable of stealing the functionality of the original model. Attackers who have difficulty obtaining triggers have no opportunity to access the model's inference results modified by the no-box scheme under the forensic state. Defense against model extraction attacks often relies on having a learnable watermark pattern within the \textit{(query, response)} data available to attackers. When an attacker successfully performs a model extraction attack on a black-box model, the attacker can easily trigger the API's forensic state, which means that the target model is far from being \textit{performance-lossless}.

However, model extraction attacks are indeed a significant threat to protecting model intellectual property in a black-box scenario. In practice, we can refer to Szylle et al.'s work~\cite{dawn}, where watermark injection operations are randomly applied to queries under the service state, sacrificing a certain degree of model performance to meet security requirements.

\subsubsection{Implementation}
\textbf{Length of trigger.} Due to the length limitation of the trigger pattern based on a secure cryptographic primitive and the small capacity of existing text watermarking, the examples given in this paper are only suitable for text generation tasks that support long text input. How to implement performance-lossless branch watermarking for short text input language models is a direction that requires further practical design considerations.

\textbf{Detection of machine-generated text.} In addition, the trigger text generated by the proposed instance is essentially machine-generated text. Although the machine-generated trigger text can maintain good semantic readability, it is not enough to achieve excellent concealment. Concealment also involves the rationality of query behavior. The query received by the generated model is usually natural text. Obviously, the machine-generated trigger text is easily detected by the existing excellent machine text detection algorithms. However, it is extremely difficult for the detector to reach 0 false alarm. A model that provides services always faces a large number of queries. When the attacker uses the machine text detector to filter the forensic attempt of the model owner, it will also lose a large number of normal user queries, which will seriously affect the piracy service. Therefore, the cost of this kind of forensic interference attack is extremely high. Of course, we still look forward to natural text hiding methods with large capacity and efficient extraction in the future to expand upon the current trigger generation methods.

\section{Conclusion}
Existing model watermarking work in black-box scenarios relies mainly on the idea of training-based backdoors, resulting in distortion of the primary task performance. In this paper, we first propose a branch backdoor-based model watermarking protocol to protect model intellectual property and propose a performance-lossless construction based the message authentication scheme. The performance-lossless property of the proposed BranchWM is reduced to the \textit{EUF-CMA} security of MAC. Then, we analyze the possible attacks during the forensic process and instantiate a secure and performance-lossless watermarking for language generation model.

\section*{Acknowledgments}
This work was supported in part by the Natural Science Foundation of China under Grant 62102386. Deep learning icon in Figure~\ref{fig:threat} and~\ref{fig:method} is made by Freepik from \href{www.flaticon.com}{www.flaticon.com}.

\ifCLASSOPTIONcaptionsoff
  \newpage
\fi

\bibliography{citation}

\newpage
\section*{Biography}
\vspace{-12cm}
\begin{IEEEbiography}[{\includegraphics[width=1in,height=1.25in,clip,keepaspectratio]{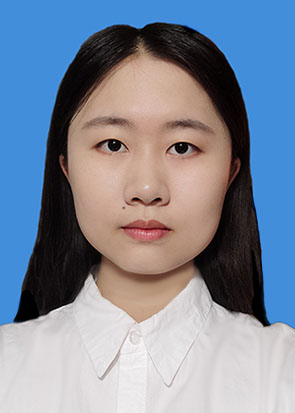}}]{Na Zhao}
received his B.S. degree in 2017 from Zhengzhou University (ZZU). Currently, she is working toward the Ph.D. degree in the University of Science and Technology of China. Her research interests include information hiding, and deep learning security.
\end{IEEEbiography}
\vspace{-12cm}
\begin{IEEEbiography}[{\includegraphics[width=1in,height=1.25in,clip,keepaspectratio]{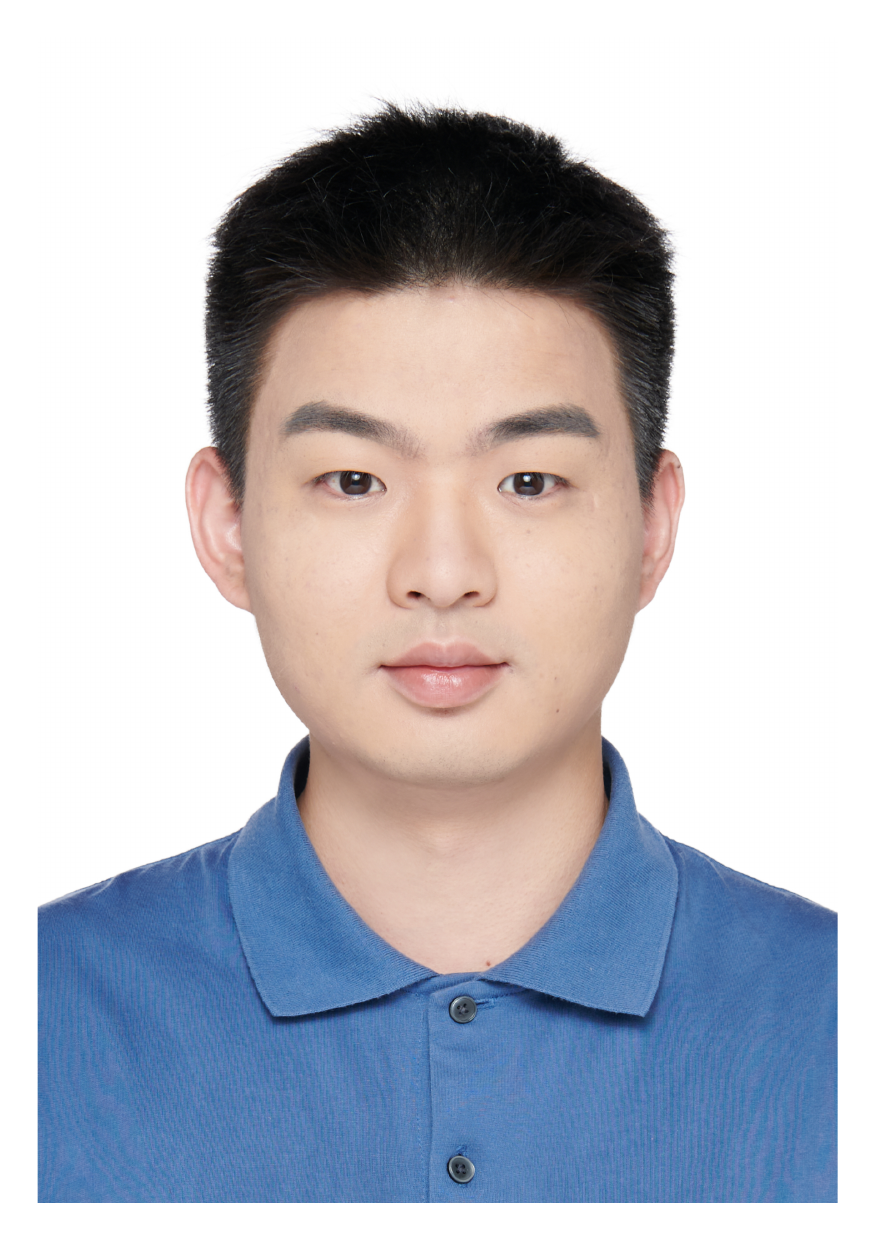}}]{Kejiang Chen}
received his B.S. degree in 2015 from Shanghai University (SHU) and a Ph.D. degree in 2020 from the University of Science and Technology of China (USTC). Currently, he is an associate researcher at the University of Science and Technology of China. His research interests include information hiding, and deep learning security.
\end{IEEEbiography}
\vspace{-12cm}
\begin{IEEEbiography}[{\includegraphics[width=1in,height=1.25in,clip,keepaspectratio]{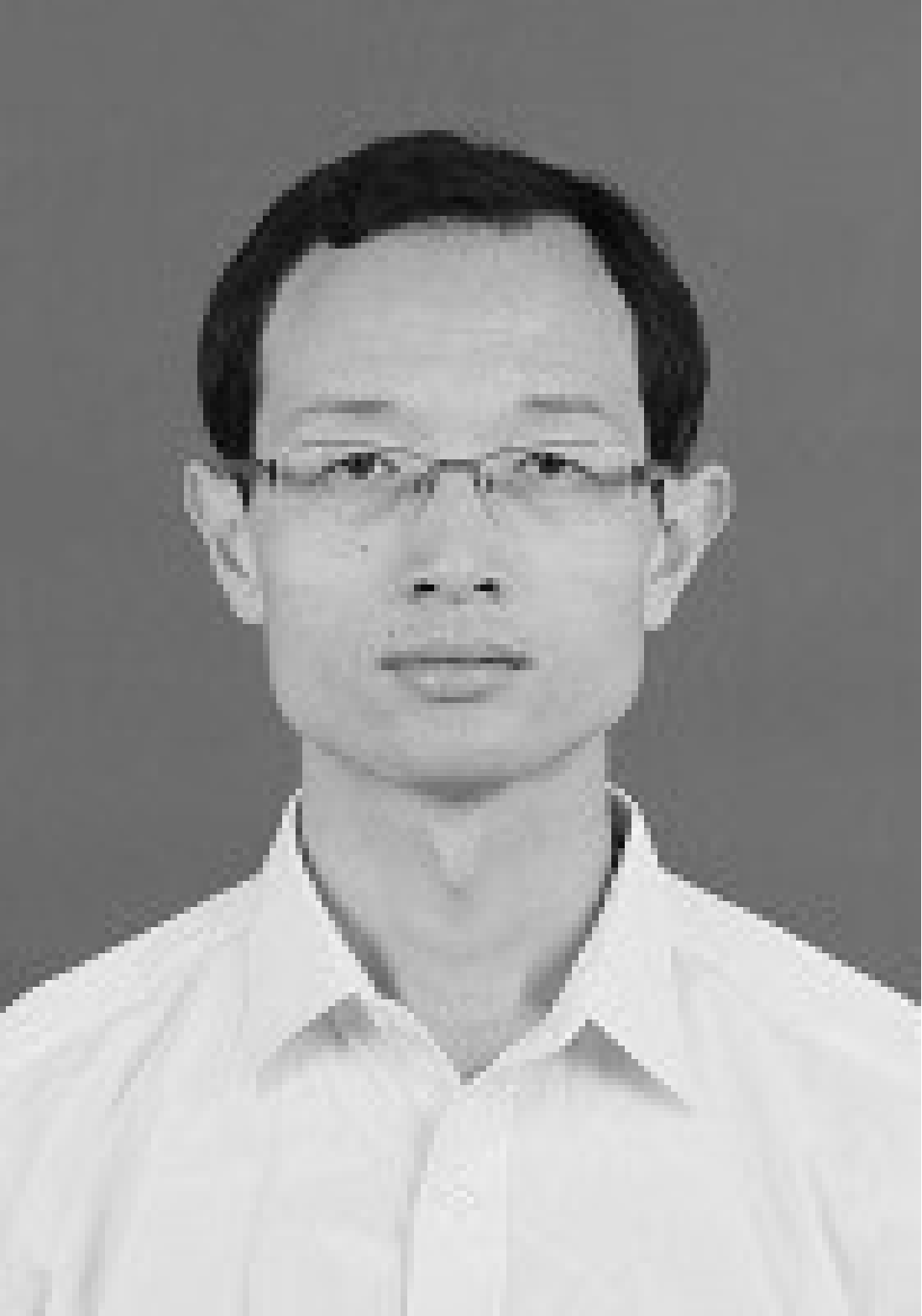}}]{Weiming Zhang}
received his M.S. degree and Ph.D. degree in 2002 and 2005, respectively, from the Zhengzhou Information Science and Technology Institute, P.R. China. Currently, he is a professor with the School of Information Science and Technology, University of Science and Technology of China. His research interests include information hiding and multimedia security.
\end{IEEEbiography}
\vspace{-12cm}
\begin{IEEEbiography}[{\includegraphics[width=1in,height=1.25in,clip,keepaspectratio]{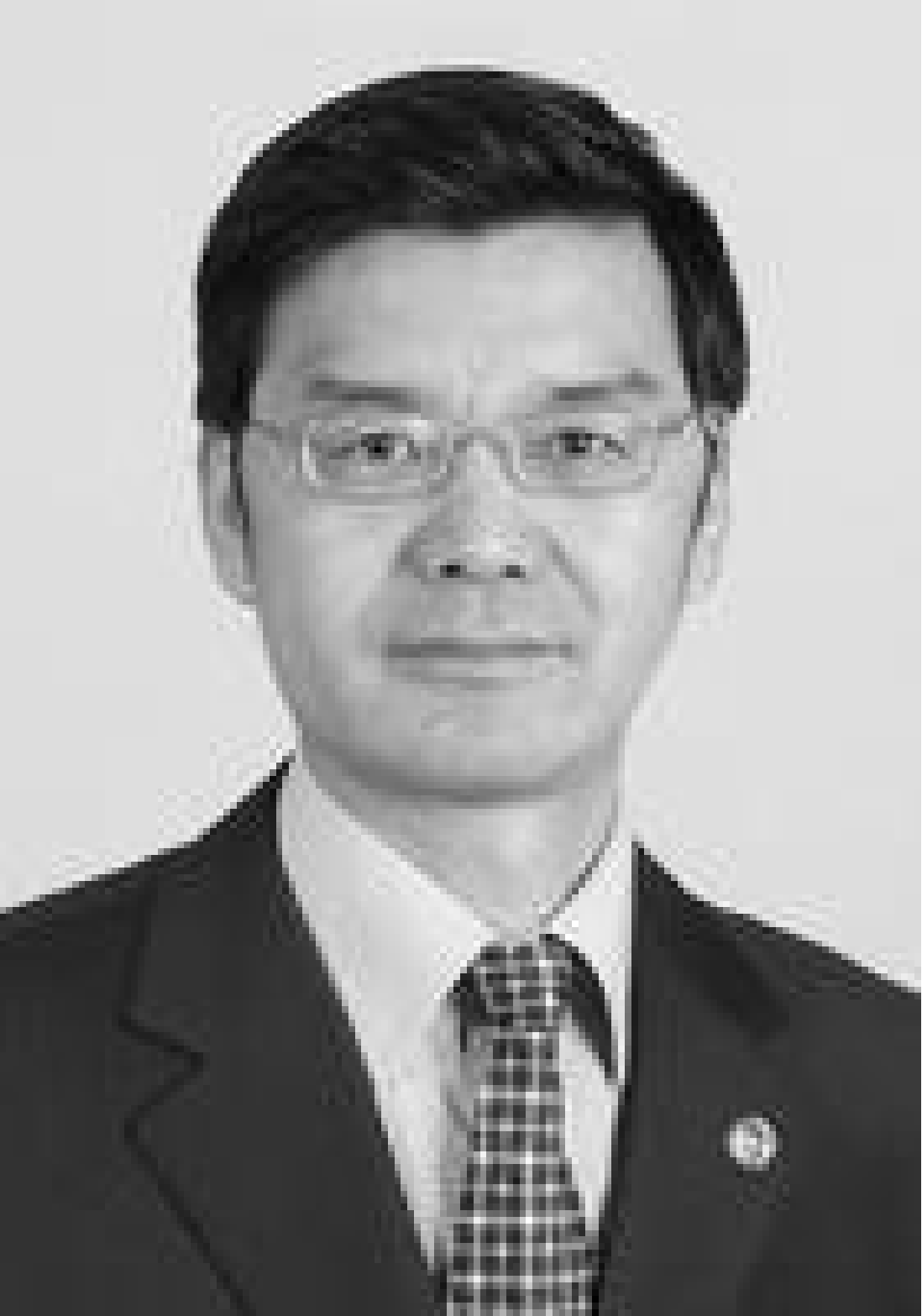}}]{Nenghai Yu}
received his B.S. degree in 1987 from Nanjing University of Posts and Telecommunications, an M.E. degree in 1992 from Tsinghua University and a Ph.D. degree in 2004 from the University of Science and Technology of China, where he is currently a professor. His research interests include multimedia security, multimedia information retrieval, video processing and information hiding.
\end{IEEEbiography}

\end{document}